\documentclass[english]{article}
\usepackage[T1]{fontenc}
\usepackage[latin9]{inputenc}
\usepackage{color}
\usepackage{amstext}
\usepackage{graphicx}
\usepackage{setspace}

\makeatletter

\DeclareRobustCommand{\greektext}{%
  \fontencoding{LGR}\selectfont\def\encodingdefault{LGR}}
\DeclareRobustCommand{\textgreek}[1]{\leavevmode{\greektext #1}}
\DeclareFontEncoding{LGR}{}{}

\makeatother

\usepackage{babel}

\begin{document}

\title{\textcolor{black}{The Tajmar effect from Quantised Inertia.}}

\author{\textcolor{black}{M.E. McCulloch}%
\thanks{\textcolor{black}{SMSE, University of Plymouth, PL4 8AA. mike.mcculloch@plymouth.ac.uk}%
}}
\maketitle
\begin{abstract}
\begin{singlespace}
\noindent \textcolor{black}{The Tajmar anomaly is an unexplained acceleration
observed by gyroscopes close to, but isolated from, rotating rings
cooled to 5K. The observed ratio between the gyroscope and ring accelerations
was $3\pm1.2\times10^{-8}$ for clockwise rotations and about half
this size for anticlockwise ones. Here, this anomaly is predicted
using a new model that assumes that the inertial mass of the gyroscope
is caused by Unruh radiation that appears as the ring and the fixed
stars accelerate relative to it, and that this radiation is subject
to a Hubble-scale Casimir effect. The model predicts that the sudden
acceleration of the ring causes a slight increase in the inertial
mass of the gyroscope, and, to conserve momentum the gyroscope must
move with the ring with an acceleration ratio of $2.67\pm0.24\times10^{-8}$
for clockwise rotations and $1.34\pm0.12\times10^{-8}$ for anticlockwise
ones, in agreement with the observations. The model predicts that
in the southern hemisphere the anomaly should be larger for anticlockwise
rotations instead, and that with a significant reduction of the mass
of the disc, the decay of the effect with vertical distance should
become measurable.}\end{singlespace}

\end{abstract}

\section*{\textcolor{black}{Keywords}}

\textcolor{black}{Cosmology, acceleration, momentum conservation,
Tajmar experiment.}

\section*{\textcolor{black}{PACS codes}}

\textcolor{black}{95.30.-k, 45.20.df, 06.30.Gv}

\textcolor{black}{\pagebreak{}}

\section{\textcolor{black}{Introduction}}

\textcolor{black}{It has been found experimentally by {[}1-3{]} that
when rings of niobium, aluminium, stainless steel and other materials
are cooled to 5K and spun, then accelerometers and laser gyroscopes,
not in frictional contact, show a small unexplained acceleration in
the same direction as the ring, with a size $3\pm1.2\times10^{-8}$
times the acceleration of the ring for clockwise rotations, and about
half that value for anticlockwise ones. This is called the Tajmar
effect and is similar to the Lense-Thirring effect (frame-dragging)
predicted by General Relativity, but is 20 orders of magnitude larger
and shows the added parity violation. The effect has not yet been
reproduced in another laboratory.}

\textcolor{black}{{[}4{]} proposed an explanation for the anomaly
that relies on a Higgs mechanism that causes the graviton to gain
mass. This theory was called the gravitometric London effect, but
it has been discredited because the inception of the Tajmar effect
(at about 25K) does not coincide with the superconducting transition
temperature, only with very low temperatures.}

\textcolor{black}{The model suggested here as an explanation for the
effect was proposed by {[}5{]}. It assumes that the inertial mass
of an object is caused by Unruh radiation which is generated by the
object's acceleration relative to other matter, and that this radiation
is subject to a Hubble-scale Casimir effect (in which longer Unruh
waves are increasingly disallowed). The model could be called Modified
Inertia due to a Hubble-scale Casimir effect (MiHsC) or perhaps Quantised
Inertia, and was tested by {[}5{]} on the Pioneer anomaly (observed
by {[}6{]}). In {[}5{]} the inertial mass ($m_{I}$) was defined as}

\textcolor{black}{\begin{equation}
m_{I}=m_{g}\left(1-\frac{\beta\pi^{2}c^{2}}{\left|a\right|\Theta}\right)\end{equation}
}

\begin{singlespace}
\noindent \textcolor{black}{where $m_{g}$ is the gravitational mass,
$\beta=0.2$ (empirically derived by Wien as part of Wien's law),
c is the speed of light, $\Theta$ is the Hubble diameter ($2.7\times10^{26}m$,
from {[}7{]} and the magnitude of the acceleration (a) in this case
was the acceleration of the Pioneer craft relative to their main attractor:
the Sun. This model predicted a small loss of inertial mass for the
Pioneer spacecraft that increased their Sunward acceleration by an
amount close to the observed Pioneer anomaly (beyond 10AU from the
Sun).}
\end{singlespace}

\noindent \textcolor{black}{Reference {[}8{]} applied MiHsC to the
unexplained velocity jumps observed in Earth flybys of interplanetary
probes (the flyby anomalies observed by {[}9{]}) and found that these
anomalies could be reproduced quite well if the acceleration in Eq.
1 was taken to be that of the spacecraft relative to all the particles
of matter in the spinning Earth. Using MiHsC and the conservation
of momentum the predicted anomalous jump for spacecraft passing by
the spinning Earth was}

\textcolor{black}{\begin{equation}
dv'=\frac{\beta\pi^{2}c^{2}}{\Theta}\left(\frac{v_{2}}{\left|a_{2}\right|}-\frac{v_{1}}{\left|a_{1}\right|}\right)\end{equation}
}

\begin{singlespace}
\noindent \textcolor{black}{where $a_{1}$ and $a_{2}$ were the average
accelerations of all the matter in the spinning Earth seen from the
point of view of the incoming ($a_{1}$) and outgoing ($a_{2}$) craft.
This formula predicted a slightly lower inertia for spacecraft close
to the Earth's spin axis, meaning that, by conservation of momentum,
their speed increases by an amount similar to the observed flyby anomalies
(except for the EPOXI flyby, which was much further from the Earth
than the others and is discussed later).}

\noindent \textcolor{black}{The Tajmar effect is similar to the flyby
anomalies, although instead of being an anomalous acceleration of
a spacecraft close to a spinning planet, it is an anomalous acceleration
of a laser gyroscope close to a cold spinning ring. Therefore, in
this paper MiHsC is applied to the Tajmar effect, but also, following
Mach's principle, in this paper the effect of the relative accelerations
of the fixed stars on inertia are also considered. A previous paper
by the author {[}10{]} applied the same theory, conserving momentum
relative to the fixed stars, and explained Tajmar's setup A and B
numerically (though the reference frame used was incorrect) and setup
B's rotation direction. However, the rotation direction in setup B
may have had a clockwise bias (Tajmar, pers. comm.). In this paper
the analysis is the same except that momentum is conserved relative
to the moving ring and the model explains very well the results from
setup A {[}3{]} (and setup B if the bias is corrected) including the
observed parity violation in setup A. As shown here, MiHsC predicts
that performing the experiment in the southern hemisphere, the gyroscope
should still follow the ring's rotation, but the parity violation
should be greater for anticlockwise ring rotation instead of for clockwise.
This is an important correction to the prediction made by {[}10{]}
where the gyroscope's rotation was predicted to be clockwise in the
north, and anticlockwise in the south.}
\end{singlespace}

\section{\textcolor{black}{Method \& Results}}

\textcolor{black}{The assumed experimental set up is that of {[}3{]}
and their set-up configuration A, and is shown in Figure 1, with a
rotating super-cooled ring of radius r. Three laser gyroscopes of
mass m are located above the ring. The fixed stars are also shown
schematically. They have a huge combined mass, but are very far away. }

\noindent \textcolor{black}{For the laser gyroscope situated just
above the ring we can assume a conservation of momentum parallel to
the ring's edge (subscript 'r')}

\textcolor{black}{\begin{equation}
m_{g1}v_{gr1}=m_{g2}v_{gr2}\end{equation}
}

\textcolor{black}{where $v_{gr}$ is the velocity of the gyroscope
(g) with respect to the ring (r), hence 'gr'. Replacing the inertial
masses with the modified inertia of {[}5{]} leaves}

\textcolor{black}{\begin{equation}
v_{gr1}\left(1-\frac{\beta\pi^{^{2}}c^{2}}{\left|a_{g1}\right|\Theta}\right)=v_{gr2}\left(1-\frac{\beta\pi^{^{2}}c^{2}}{\left|a_{g2}\right|\Theta}\right)\end{equation}
}

\textcolor{black}{and rearranging}

\textcolor{black}{\begin{equation}
v_{gr2}-v_{gr1}=\frac{\beta\pi^{^{2}}c^{2}}{\Theta}\left(\frac{v_{gr2}}{\left|a_{g2}\right|}-\frac{v_{gr1}}{\left|a_{g1}\right|}\right)\end{equation}
}

\noindent \textcolor{black}{This is similar to Eq. 2, which was derived
from MiHsC for the flyby anomalies. For this new case, the initial
and final accelerations ($a_{g1}$ and $a_{g2}$) of the gyroscope
with respect to all the surrounding masses now need to be defined.
First we assume that because of cooling the temperature-dependent
acceleration of nearby atoms is small. We can say that the acceleration
relative to the atoms in the Earth is zero since the experiment is
solidly fixed to the Earth. So, before the ring accelerates the gyroscope
sees only an acceleration of the fixed stars since it is on the spinning
Earth. These are far away, but their combined mass is huge. The rotational
acceleration with respect to the fixed stars ($a_{s}$) of an object
fixed to the Earth at the latitude of Seibersdorf in Austria where
the experiment was performed (at 48$^{\text{0}}N$) is the same as
the Coriolis acceleration: fv, where $f\sim0.0001s^{-1}$ in mid-latitudes,
and v, the spin velocity of the Earth at this latitude is 311 m/s,
so $a_{s}=0.0311m/s^{2}$. To this should be added the acceleration
due to the Earth's orbit around the Sun (so we have: $a_{s}=(0.0311+0.006)m/s^{2})$.
The acceleration due to the Sun's orbit around the galaxy is far smaller
and can be neglected. So in the above formula $a_{g1}=0.0371m/s^{2}$.}

\noindent \textcolor{black}{The sudden acceleration of the Tajmar
ring causes an acceleration of $a_{R}=33\, rad/s^{2}=2.5\, m/s^{2}$
(since the radial position of the gyroscope was 0.075 m). Therefore
$a_{g2}=fn(a_{s},a_{R})$. However, to find the average acceleration
we have to consider the relative importance of the fixed stars and
the ring for determining the inertia of the gyroscope. As in {[}10{]}
we will assume here that the importance of an object for the inertia
of another one is equivalent to its gravitational importance, which
is proportional to its mass over the distance squared. The details
of this assumption do not effect the final result as we will see.
Therefore $a_{g2}$ is}

\textcolor{black}{\begin{equation}
a_{g2}=\frac{\frac{m_{s}}{r_{s}^{2}}a_{gs}+\frac{m_{R}}{r_{R}^{2}}a_{gr}}{\frac{m_{s}}{r_{s}^{2}}+\frac{m_{R}}{r_{R}^{2}}}\end{equation}
}

\noindent \textcolor{black}{where $m_{s}$ is the mass of all the
fixed stars, and $r_{s}$ is their mean distance away and $m_{R}$
is the mass of the ring and $r_{R}$ is its distance away. Using Eq.
6 in Eq. 5 gives}

\textcolor{black}{\begin{equation}
v_{gr2}-v_{gr1}=\frac{\beta\pi^{2}c^{2}}{\Theta}\left(\frac{v_{gr2}}{\left|\frac{\frac{m_{s}}{r_{s}^{2}}a_{gs2}+\frac{m_{R}}{r_{R}^{2}}a_{gr2}}{\frac{m_{s}}{r_{s}^{2}}+\frac{m_{R}}{r_{R}^{2}}}\right|}-\frac{v_{gr1}}{\left|a_{gs1}\right|}\right)\end{equation}
}

\begin{singlespace}
\noindent \begin{flushleft}
\textcolor{black}{These values were approximated as a total stellar
mass of $m_{s}\sim2.4\times10^{52}kg$ from {[}11{]}, at a distance
of $r_{s}\sim2.7\times10^{26}m$ ($r_{s}=2c/H$, derived from the
Hubble constant, H, from {[}7{]}), and a ring mass of $m_{R}\sim0.336kg$
(stainless steel has a density of about 8000 kg/m$^{\text{3}}$, and
the ring had a circumference of 2{*}\textgreek{p}{*}0.075, a height
of 0.015 m and a width of 0.006 m) and a ring distance of $r_{R}\sim0.0533m$
(the estimated vertical distance from the centre of the lower gyroscope
to the ring). Using these values we have}
\par\end{flushleft}
\end{singlespace}

\textcolor{black}{\begin{equation}
v_{gr2}-v_{gr1}=\frac{\beta\pi^{2}c^{2}}{\Theta}\left(\frac{v_{gr2}}{\left|\frac{0.33a_{gs2}+118a_{gr2}}{118}\right|}-\frac{v_{gr1}}{\left|a_{gs1}\right|}\right)\end{equation}
}

\textcolor{black}{We can therefore neglect $a_{gs2}$ in the denominator
of the first term in brackets (this simplification similarily applies
to the aluminium and niobium rings) to give}

\textcolor{black}{\begin{equation}
v_{gr2}-v_{gr1}\sim\frac{\beta\pi^{2}c^{2}}{\Theta}\left(\frac{v_{gr2}}{\left|a_{gr2}\right|}-\frac{v_{gr1}}{\left|a_{gs1}\right|}\right)\end{equation}
}

\textcolor{black}{Now, since the ring is spinning fast at time 2 the
first term in the brackets is small (since $a_{gr2}=v_{gr2}^{2}/r$)
so we can say}

\textcolor{black}{\begin{equation}
dv'\sim\frac{\beta\pi^{2}c^{2}}{\Theta}\left(-\frac{v_{gr1}}{\left|a_{gs1}\right|}\right)\end{equation}
}

\textcolor{black}{Differentiating with respect to time to find the
resulting anomalous acceleration and neglecting changes in $a_{gs}$
(for now):}

\textcolor{black}{\begin{equation}
da'\sim-\frac{\beta\pi^{2}c^{2}}{\Theta}\frac{a_{gr}}{\left|a_{gs}\right|}\sim\frac{-6.6\times10^{-10}}{0.0371}a_{gr}\end{equation}
}

\textcolor{black}{Changing $a_{gr}$ to $a_{rg}$ (which involves
a sign change) we get}

\textcolor{black}{\begin{equation}
da'\sim\frac{\beta\pi^{2}c^{2}}{\Theta}\frac{a_{rg}}{\left|a_{gs}\right|}\sim\frac{6.6\times10^{-10}}{0.0371}a_{rg}\sim1.78\pm0.16\times10^{-8}a_{rg}\end{equation}
}

\textcolor{black}{The $a_{rg}$ is the rotational acceleration of
the ring with respect to the gyroscope so Eq. 12 implies that the
anomalous rotational acceleration of the gyroscope will be in the
same direction as the ring, as observed by {[}3{]} for set-up A and
the predicted acceleration ratio is $1.78\pm0.16\times10^{-8}$ in
agreement with the observed value which was $3\pm1.2\times10^{-8}$.
The predicted error of 0.16 was derived by assuming a 9\% error in
the Hubble constant (and therefore the estimated Hubble diameter)
following {[}7{]}. Of the two ratios compared here, one is a observed
velocity ratio and one is a predicted acceleration ratio. These are
the same because the acceleration of the gyro (da') is not an spin-up
of the gyro's field coil itself but an acceleration of the gyro along
the ring's edge, in the same radius circle as the ring. Therefore
in the formula $a=v^{2}/r$, r is the same in both cases, so the v
and acceleration ratios are the same.}

\textcolor{black}{At the top speed of the ring the anomalous gyroscope
output (its spin) was equal to one third of the Earth's rotation,
according to {[}3{]}. The $a_{gs}$ in equations 10-12 then also depends
on the gyroscope's rotation. If the ring (and then the gyro) rotate
clockwise (anticyclonically) then this is counter to the Earth's rotation
and so the original acceleration of the gyro with respect to the fixed
stars: $a_{gs}$ decreases by one third, therefore $1/a_{gs}$ increases
by 1.5 times, and so the predicted anomalous signal from Eq. 12 for
clockwise rotations increases from $1.78\pm0.16\times10^{-8}$ to
$2.67\pm0.24\times10^{-8}$ (the observation was $3\pm1.2\times10^{-8}$)
. If the ring and gyroscope rotate anticlockwise then this adds to
the Earth's own spin, it increases $a_{gs}$ by one third, so $1/a_{gs}$
decreases to 75\% of its original value and the predicted signal decreases
to $1.34\pm0.12\times10^{-8}$ and this agrees with the observed anticlockwise
value which was about half of the clockwise value: $1.5\pm1.2\times10^{-8}$,
according to {[}3{]} (Fig. 2).}

\section{\textcolor{black}{Discussion}}

\textcolor{black}{In the model used here (MiHsC) the inertial mass
of the gyroscope is assumed to be determined by Unruh radiation that
appears as it accelerates with respect to every other mass in the
universe. The Unruh radiation is also subject to a Hubble-scale Casimir
effect. Before its surroundings are cooled the gyroscope sees large
accelerations due to the vibration of nearby atoms, it is surrounded
by Unruh radiation of short wavelengths, MiHsC has only a small effect,
and the inertial mass of the gyroscope is close to its gravitational
mass. The nearby atomic accelerations reduce when the surroundings
(the cryostat) are cooled, so that the inertia of the gyroscope is
more sensitive to the accelerations of the fixed stars (it is on the
rotating Earth). This is a small acceleration, so the Unruh waves
it sees are long and a greater proportion are disallowed by MiHsC,
so the gyroscope's inertial mass (from Eq. 1) falls very slightly
below its gravitational mass. In this case it looses $2\times10^{-8}kg$
for every kilogram of mass. However, when the nearby ring accelerates,
the gyroscope suddenly sees the higher accelerations of the ring,
the Unruh waves shorten, fewer are disallowed, and its inertial mass
increases again following Eq. 1. The important point is that to conserve
momentum with respect to the ring, the heavier gyroscope must accelerate
in the same direction as the ring. A further complication is that
when the gyroscope does start to spin with the ring, it changes its
original acceleration with respect to the fixed stars, producing the
parity violation.}

\textcolor{black}{MiHsC (Eq. 1) violates the equivalence principle,
but not in a way that could be detected by the usual torsion balance
experiment. These experiments measure the differential attraction
of two balls on a cross bar suspended on wire, towards distant masses
by detecting tiny twists in the wire (eg: {[}12{]}). With MiHsC these
two balls would have equal accelerations with respect to the distant
masses (being rigidly connected) so their inertial masses would be
modified equally by MiHsC, and there will be no twist in the wire,
and no apparent violation of equivalance.}

\textcolor{black}{Eq. 7 predicts a decay in the MiHsC effect with
distance. For this experiment, setup A, at 1m, 5m, 20m and 56m away
the effect is predicted by Eq. 7 to be 0.03\%, 0.8\%, 11.5\% and 50\%
smaller. The decay at 20m may be detectable, but cryostats this long
are difficult to find. The effect should decay more quickly outside
the cryostat, tending to a decay proportional to one over distance
squared due to nearby thermal accelerations, but this needs to be
studied in more detail. (See {[}10{]} for a more detailed discussion
of the decay). Calculations using Eq. 7 show that a change in the
ring velocity does not change the measured gyro/ring ratio at the
lower gyroscope (0.0553m away) much, but it does change the decay
of the effect with distance. For example, a }\textit{\textcolor{black}{reduction}}\textcolor{black}{
in ring acceleration by a factor of 10,000 (a reduction of velocity
by a factor of 100) will increase the decay of the MiHsC effect with
distance, so that the upper gyroscope 0.2283m away would see a decrease
of 13.6\%, but the gyroscope may not be able to measure the lower
velocities. A better way to achieve the same result would be to reduce
the ring's mass. If this was reduced by a factor of 10,000 the gyro/ring
acceleration ratio would remain the same at the lower gyroscope 0.0553m
away, but the gyroscope 0.2283m away should now see a 13.6\% drop
in the effect. Tajmar's group could try this test with their existing
equipment.}

\textcolor{black}{As discussed in the introduction, {[}8{]} applied
MiHsC quite successfully to predicting the flyby anomalies. Subsequent
data obtained from J.Campbell of NASA (at a flyby workshop organised
by the International Space Science Institute, ISSI, in Bern, Switzerland)
shows that the flyby anomaly for the EPOXI spacecraft was zero, whereas
MiHsC predicts a large anomaly. A difference with the EPOXI flyby
was that it has a periapse radius of 49,835 km (J. Campbell, pers.
comm.) from the Earth whereas the other flybys were closer, typically
7,000 km away. As stated by {[}10{]}, in MiHsC the effect of one body
on the inertia of another decays as its mass over its distance squared
and therefore the sensitivity of the EPOXI craft's inertia to the
spin of the Earth should be lower, and its sensitivity to other (more
constant) relative accelerations from other Solar system bodies should
be more important. The flyby anomalies can be predicted more successfully,
if the accelerations within the Sun are assumed to be $1m/s^{2}$.
It is not known whether this value is realistic.}

\begin{singlespace}
\noindent \textcolor{black}{A previous paper by the author applied
the same theory {[}10{]}, but it is now thought that the reference
frame used was wrong, so that the proposed test (an exact mirror-image
result in the southern hemisphere) was wrong. The new prediction is
that performing the experiment in the southern hemisphere the gyroscope
should still follow the ring's rotational direction, but the anomalous
signal should be greater for anticlockwise ring rotations instead
of for clockwise ones in the north.}
\end{singlespace}

\section{\textcolor{black}{Conclusions}}

\textcolor{black}{The anomalous clockwise accelerations observed by
laser gyroscopes close to rotating super-cooled rings (see {[}3{]},
for set-up A) can be predicted by a theory (MiHsC) that assumes that
the inertial mass of the gyroscope is caused by Unruh radiation that
appears because of its mutual acceleration with the fixed stars and
the spinning ring, and that this radiation is subject to a Hubble-scale
Casimir effect.}

\textcolor{black}{The parity violation observed by {[}3{]} for setup
A can also be explained by MiHsC as a secondary effect brought on
by the movement of the gyro relative to the fixed stars. This spin
with respect to the stars, changes the gyro's inertia and its anomalous
acceleration depending on the spin direction, in good agreement with
the observations.}

\textcolor{black}{It is proposed that the validity of MiHsC in this
case could be tested by reducing the mass of the ring in {[}3{]} setup
A by a factor of 10,000 or more, and looking for a measurable decay
of the effect with vertical distance.}

\section*{\textcolor{black}{Acknowledgements}}

\textcolor{black}{Thanks to M. Tajmar, J. Campbell (and an ISSI flyby
workshop) for information, an anonymous reviewer for advice, and B.
Kim for support and encouragement.}

\section*{\textcolor{black}{References}}

\textcolor{black}{{[}1{]} Tajmar, M., F. Plesescu, B. Seifert, and
K. Marhold, 2007. Measurement of gravitomagnetic and acceleration
fields around rotating superconductors. Proceedings of the STAIF-2007
Conference, AIP Conference Proceedings, Vol. 880, pp. 1071-1082.}

\textcolor{black}{{[}2{]} Tajmar, M., F. Plesescu, B. Seifert, R.
Schnitzer, and I. Vasiljevich, 2008. Investigation of frame-dragging-like
signals from spinning superconductors sing laser gyrocopes. AIP conference
Proceedings, Vol. 69, pp 1080-1090.}

\textcolor{black}{{[}3{]} Tajmar, M., F. Plesescu and B. Seifert,
2009. Anomalous fiber optic gyroscope signals observed above spinning
rings at low temperature. J. Phys. Conf. Ser., 150, 032101, 2009.}

\textcolor{black}{{[}4{]} de Matos, C.J. and M. Tajmar, 2005. Gravitometric
London moment and the graviton mass inside a superconductor, Physica
C, 432, 167-172.}

\textcolor{black}{{[}5{]} McCulloch, M.E., 2007. Modelling the Pioneer
anomaly as modified inertia. }\textit{\textcolor{black}{MNRAS}}\textcolor{black}{,
376, 338-342.}

\textcolor{black}{{[}6{]} Anderson, J.D, P.A. Laing, E.L. Lau, A.S.
Liu, M.M. Nieto and S.G. Turyshev, 1998. Phys. Rev. Lett., 81, 2858.}

\textcolor{black}{{[}7{]} Freedman, W.L., 2001. Final results fom
the Hubble space telescope key project to measure the Hubble constant.
ApJ, 553, 47-72.}

\textcolor{black}{{[}8{]} McCulloch, M.E., 2008. Modelling the flyby
anomalies using a modification of inertia. }\textit{\textcolor{black}{MNRAS-letters}}\textcolor{black}{,
389, L57-60.}

\textcolor{black}{{[}9{]} Anderson, J.D., Campbell, J.K., Ekelund,
J.E., Ellis J., Jordan J.F., 2008. Anomalous orbital energy changes
observed during spacecraft flybys of Earth. }\textit{\textcolor{black}{Phys.
Rev. Lett.}}\textcolor{black}{, 100, 091102.}

\textcolor{black}{{[}10{]} McCulloch, M.E., 2010. Can the Tajmar effect
be explained using a modification of inertia. }\textit{\textcolor{black}{Europhys.
Lett.}}\textcolor{black}{, 89, 19001.}

\textcolor{black}{{[}11{]} Funkhouser, S., 2006. The large number
coincidence: the cosmic coincidence and the critical acceleration.
Proc. R. Soc A., vol 462, no 2076, p3657-3661.}

\textcolor{black}{{[}12{]} Gundlach, J.H., S. Schlamminger, C.D. Spitzer,
K.-Y. Choi, 2007. Laboratory test of Newton's second law for small
accelerations. }\textit{\textcolor{black}{PRL}}\textcolor{black}{,
98, 150801.}

\section*{\textcolor{black}{Figures}}

\textcolor{black}{\includegraphics[scale=0.8]{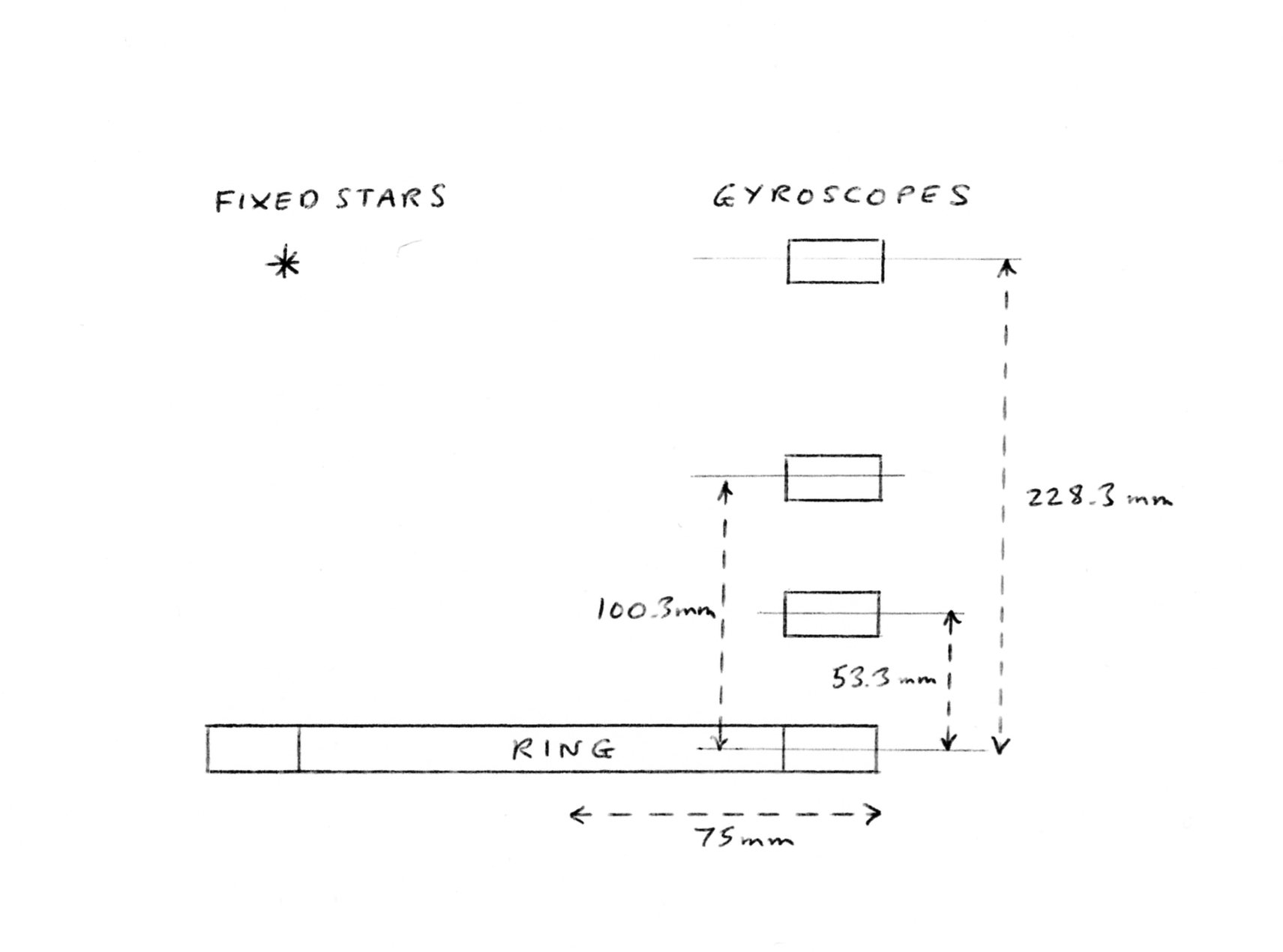}}

\textcolor{black}{Figure 1. Schematic showing the experiment of {[}3{]},
their setup A. The ring, the lower, middle and upper laser gyroscopes,
some dimensions and the fixed stars are shown.}
\end{document}